# A Longitudinal Study of Social Media Privacy Behavior

Andrew W. Boyd

*Seidenberg School of CSIS, Pace University, White Plains, NY 10606, USA*

*Abstract* - **Existing constructs for privacy concerns and behaviors do not adequately model deviations between user attitudes and behaviors. Although a number of studies have examined supposed deviations from rationality by online users, true explanations for these behaviors may lie in factors not previously addressed in privacy concern constructs. In particular, privacy attitudes and behavioral changes over time have not been examined within the context of an empirical study. This paper presents the results of an Agile, sprint-based longitudinal study of Social Media users conducted over a two year period between April of 2009 and March of 2011. This study combined concepts drawn from Privacy Regulation Theory with the constructs of the Internet Users' Information and Privacy Concern model to create a series of online surveys that examined changes of Social Media privacy attitudes and self-reported behaviors over time. The main findings of this study are that, over a two year period between 2009 and 2011, respondents' privacy concerns and distrust of Social Media Sites increased significantly, while their disclosure of personal information and willingness to connect with new online friends decreased significantly. Further qualitative interviews of selected respondents identified these changes as emblematic of users developing ad-hoc risk mitigation strategies to address privacy threats.**

## 1. Introduction

Online privacy risk has emerged as one of the largest threats facing Internet users [5-7]. In the past five years, deficiencies in Social Networking Site privacy management have come under particular scrutiny from academics and have become the source of thousands of leading articles within the popular press. Consequently, a great deal of fear, uncertainty and doubt surround discourse tied to online technologies and few positive solutions have been suggested. The current situation has been described as a "Privacy Train Wreck" [6] and no clear remediation strategy has yet been proposed.

Social Networking Sites provide excellent study domains for online privacy behavior due to the manner in which they merge technology with sensitive privacy data. These sites, by their very nature, encourage users to share personal and professional information and within the social sciences [8], computer science [9], and the popular media [10], there is growing concern over how Social Networking Sites collect and use personal information and how this information is shared among site users. In particular, revenue generating business models for Social Networking Sites, content publishers, search engines, and web analytics aggregators represent significant threats to personal privacy [10, 11].

This research adds to the existing body of knowledge within security and privacy studies, particularly within the rapidly evolving field of social media studies. Given the high level of privacy risk presented by Social Networking Sites, a closer examination of user attitudes and behaviors is both timely and greatly needed. Publications from 2006 [4], 2008 [26], 2009 [11] and as recently as November 2010 [27] have called for a longitudinal study of this nature, and it is probable that the only reason a longitudinal privacy study has not yet been produced is the relative infancy of social media and the labor investment required to track these phenomena over an extended period of time.

This research addresses two of the main questions within the field of privacy and security studies: what variables influence online privacy attitudes and behaviors, and how do these variables evolve over time? At a more nuanced level, this study interrogates the fluidity of personal privacy boundaries and bi-directional correlations among the above variables that operate along those privacy disclosure boundaries. As part of the Agile, iterative and evolutionary nature of this study; the Internet Users' Information Privacy Concern model (IUIPC) [12], an empirically validated survey instrument, was selected as a scientifically rigorous counterweight to the agility of sprint-based hypothesis testing and revision.

The first research question is investigated by extending the IUIPC model for Social Networking Site applicability and by extending the co-variates tested within the model. The second question is addressed by iteratively applying this extended model over a period of nineteen months between April of 2009 and December of 2010 and testing a set of hypotheses that explain changes in attitudes and behaviors over this period.

Additional research questions are addressed with the course of this study. These questions include the role of demographics, privacy knowledge and expertise, ethnography, and contextual circumstance in influencing online privacy attitudes and behaviors.

## 2. Hypothesis Development

After surveying the bibliography of work in this area, and having determined that gaps between privacy attitudes and stated privacy behaviours are not adequately explained by existing constructs, it was determined that the Internet User Information Privacy Concern model provided the best 'core' set of questions for this study's requirements [1][2][3][13][15]. IUIPC has been referenced and implemented in over 14 studies, and has achieved relatively wide acceptance.

However, I also wished to establish a baseline for the stated behaviours of internet users based on a range of demographic co-variants. This would enable the testing of demographic co-variants for their impact upon privacy attitudes and stated behaviours towards social media. Accordingly, a series of questions was constructed to specifically target personal information disclosure within the context of social networking communities. By iteratively applying a common set of questions across multiple applications of the survey instrument, it became possible to test longitudinal changes in privacy attitudes and behaviors.

These hypotheses and findings are outlined in the following summary chart.

**Table 1: Longitudinal Study Hypotheses**

| Number | Hypothesis | Finding |
|--------|------------|---------|
| H1 | Social media privacy concerns are positively correlated to time. | Supported |
| H2 | Distrust of Social Networking Sites is positively correlated to time. | Supported |
| H3 | Risk Perception for privacy disclosure is positively correlated to time. | Supported |
| H4 | Privacy disclosure to Social Networking Sites is negatively correlated to time. | Supported |

A more detailed description of these hypotheses and an examination of how they vary over time appears in the 'Findings' section of this paper.

**IUIPC model**

The IUIPC model draws upon Social Contract theory to present a theoretical framework consisting of multidimensional first and second order elements, as well as a series of demographic covariates.

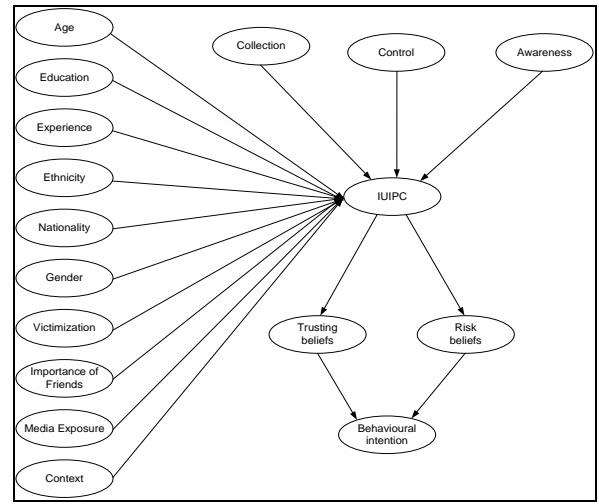

**Figure 1: Longitudinal Study extension of IUIPC**

The IUIPC construct states that individual attitudes towards the collection and control of personal information and awareness of information privacy practices constitute a user's IUIPC profile [1]. This individual IUIPC profile influences trusting beliefs and risk beliefs, which in turn have an impact upon behavioral intent.

Demographic co-variants are also related to individual IUIPC profiles [1][9]. A limited number of demographic co-variants were included in the initial version of IUIPC proposed by Malhorta et al, and this study expands that list to include social media specific co-variants. This study's hypotheses posit that specific co-variants are correlated to specific privacy attitudes.

## 3. Methodology

A survey was developed based upon the IUIPC survey questions. Our survey adapted these questions for a social media context, and added a series of questions to establish demographic co-variates for each respondent. Demographic questions utilized the appropriate response categories for questions about gender, age, education, ethnography and nationality, while IUIPC questions were scored on a seven-point Likert Scale.

The survey was administered four times through SurveyMonkey, an online survey engine, over four iterations during the months of April 2009, November and December 2009; April and May 2010 and December 2010. Follow-up qualitative interviews were conducted with a representative subset of respondents during the months of February and March of 2011. Each iteration solicited survey participants drawn from two 'solicitation pools' throughout the period of the study.

For each iteration the SurveyMonkey link was distributed to the approximately 150 Canadian longitudinal panel respondents via broadcast emails, and to an ad-hoc distribution list of approximately 150 Pace University students and faculty. As each iteration's hypotheses were proven or disproven by statistical analysis, the study model was refined, new hypotheses generated, and then tested in the next iteration of the study. Thus, this study clarified research questions by progressively focusing in upon specific key elements of online privacy behavior in an iterative, empirically based manner. While the two respondent groups shared many similar demographic and response attributes, particularly in the areas of changing attitudes over time, they are fundamentally different groups. The Canadian respondent group membership constituted a longitudinal panel study, while the Pace University respondent group comprised an ad-hoc collection of respondents whose membership varied among iterations of the survey.

Although these two groups displayed similar changes in attitudes and behavior over time, only the Canadian panel is suitable for inclusion in a scientifically rigorous longitudinal study. However, qualitative interviews revealed that Canadian and American respondents employ similar privacy strategies and it can be suggested that a common effect is at work in both groups.

The survey asked a series of baseline questions to establish the demographic co-variates represented in the extended IUIPC diagram below, and to determine specific aspects of stated user behavior expressed by types of information listed in Social Network Profiles and frequency of social network usage. Users then completed a series of questions asking what type of information they make available on the Facebook, MySpace, LinkedIn, Twitter and Flixster Social Networking Sites.

The core IUIPC segment of the survey consisted of three questions about individual privacy concerns for control of personal information on Social Networking Sites, four questions about individual privacy concerns about the collection of personal information by Social Networking Sites, and three questions regarding individual privacy concerns about awareness of information privacy practices. These questions were then followed by two hypothetical scenarios that evaluated if users would accept a Social Networking Site friend request from a known individual vs. a friend request from an unknown individual.

Data were collected from the survey engine in the form of individual survey responses, formatted in Excel Spreadsheets, and subjected to statistical analysis. Data analysis and hypothesis testing using the Chi-Square and ANOVA statistical methodologies were used to test statistical significance of results against hypotheses for each iteration of this study, and the results of these analyses were used as inputs to elaborate and modify the conceptual model presented in this study.

## 4. Findings

The following findings were derived from the four iterations of the study instrument over a period of two years between April of 2009 and March of 2011.

### 4.1 Demographics

The Canadian longitudinal panel remained consistent throughout the course of the study, with an average of approximately 90% of invited respondents participating in each iteration. Although aggregate Canadian longitudinal panel survey response varied slightly by iteration, a very high percentage of panel respondents was tracked throughout this longitudinal study. The demographic profile data presented below also shows significant consistency across iterations of the study. High response rates and consistent demographics underscore the validity of the membership for this longitudinal panel study.

**Gender Ratios**
Gender ratios remained relatively consistent across iterations, with a slight majority of male respondents.

**Table 2 – Canadian Panel Gender Ratios**

| Gender | Iteration 1 | Iteration 2 | Iteration 3 | Iteration 4 |
|--------|-------------|-------------|-------------|-------------|
| Male   | 78 | 76 | 81 | 77 |
| Female | 55 | 58 | 58 | 59 |
| **Total** | **133** | **134** | **139** | **136** |

**Age Ratios**
Age ratios for the Canadian longitudinal panel also remained relatively consistent across iterations, and combined ratios are depicted in the figure below.

**Table 3 – Canadian Panel Age Ratios**

| Age | Iteration 1 | Iteration 2 | Iteration 3 | Iteration 4 |
|-----|-------------|-------------|-------------|-------------|
| 15 – 24 | 23 | 23 | 26 | 21 |
| 25 – 34 | 31 | 32 | 34 | 33 |
| 35 – 44 | 35 | 35 | 38 | 37 |
| 45 – 54 | 31 | 31 | 32 | 34 |
| 55 - 64 | 11 | 11 | 8 | 9 |
| Over 65 | 2 | 2 | 1 | 2 |
| **Total** | 133 | 134 | 139 | 136 |

**Education Levels**
Canadian longitudinal panel participants were fairly well educated; at least 80% of respondents completed high school, and approximately 30% of respondents had post-graduate degrees. Ratios for all respondents across all four iteration of the study appear in the figure below.

**Table 4 – Canadian Panel Education Levels**

| Education | Iteration 1 | Iteration 2 | Iteration 3 | Iteration 4 |
|-----------|-------------|-------------|-------------|-------------|
| Some High | 7 | 7 | 8 | 6 |
| High School Grad | 17 | 18 | 19 | 17 |
| Some College | 28 | 29 | 29 | 27 |
| College Graduate | 42 | 41 | 44 | 45 |
| Masters Degree | 19 | 21 | 20 | 22 |
| Professional Degree | 18 | 17 | 18 | 17 |
| Doctorate | 2 | 1 | 2 | 1 |
| Total | 133 | 134 | 139 | 136 |

**Ethnographic Ratios**
Study respondents were a relatively diverse group. Respondents of Western European origin constituted approximately 50% of the Canadian longitudinal panel respondents in all iterations of the survey.

**Table 5 – Canadian Panel Ethnicity**

| Age | Iteration 1 | Iteration 2 | Iteration 3 | Iteration 4 |
|-----|-------------|-------------|-------------|-------------|
| African-American | 8 | 6 | 7 | 7 |
| Western European | 66 | 70 | 68 | 72 |
| East Asian or Pacific Islander | 16 | 22 | 18 | 15 |
| South Asian | 19 | 18 | 21 | 20 |
| Hispanic | 7 | 6 | 8 | 7 |
| Middle Eastern | 11 | 12 | 9 | 9 |
| Native American | 0 | 0 | 0 | 1 |
| Other | 6 | 0 | 8 | 5 |
| **Total** | 133 | 134 | 139 | 136 |

## 4.2 Hypothesis Testing

Statistical analyses for major study hypotheses appear in the following section. For all hypotheses, thresholds of significance were set at 0.01.

### Social Media Privacy Concerns

H1: *Social media privacy concerns are positively correlated to time.* Privacy concerns of the Canadian longitudinal panel increased significantly during the period of this study. At this point, the results are mostly 'pinned' against scalar maximums and further movement is unlikely. In short, users are about as concerned as they possibly can be.

When depicted visually, this information displays a clear trend of increasing privacy concerns on the part of consumers.

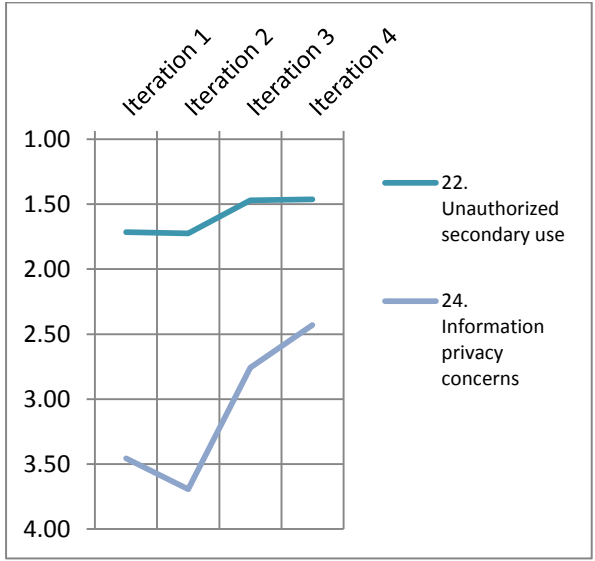

**Figure 2 – Privacy Concerns 2009-2011**

The figure above delineates the Canadian Longitudinal Panel's increase in privacy concerns over the study period. The y-axis denotes the means of responses on a 7 point Likert Scale where 1= "Strongly Agree" and 7="Strongly Disagree". A lower number indicates a higher level of privacy concern. With a Chi-Square probability of 0.001 for increases in respondent concerns about unauthorized secondary use of personal and general information privacy concerns between Iteration 1 and Iteration 4, this shift was highly statistically significant.

### Distrust of Social Networking Sites

H2: *Distrust of Social Networking Sites is positively correlated to time.* Distrust of Social Networking Sites by the Canadian Longitudinal Panel has increased significantly during the period of this study. With a Chi-Square probability of 0.0004 between Iteration 1 and Iteration 4, this shift was highly statistically significant.

### Risk Perception

H3: *Risk Perception for privacy disclosure is positively correlated to time.* Risk perception for privacy disclosure by the Canadian Longitudinal Panel increased significantly during the period of this study. With a Chi-Square probability of 0.0008 between Iteration 1 and Iteration 4, this shift was highly statistically significant.

### Privacy Disclosure

H4: *Privacy disclosure to Social Networking Sites is negatively correlated to time.* Privacy disclosure to Social Networking Sites by the Canadian Longitudinal Panel decreased significantly during the period of this study. With a Chi-Square probability of 0.0002 between Iteration 1 and Iteration 4, this shift was highly statistically significant.

The following figure shows changes in the above hypotheses during the period of this study. The y-axis denotes the means of responses on a 7 point Likert Scale where 1="Strongly Agree" and 7="Strongly Disagree". In this chart, a higher number indicates greater distrust, perception of greater risk and greater refusal to disclose.

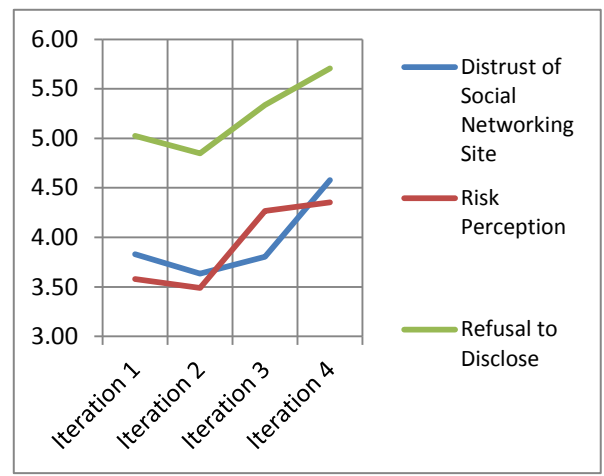

**Figure 3 – Disclosure Behaviors 2009 – 2011**

This study found that social media privacy attitudes and behaviors varied significantly over time, with the intent to disclose personal information and amount of personal information disclosed negatively correlated to time for the Canadian longitudinal panel, and the sense of immediate privacy threat risk and distrust of Social Networking Sites positively correlated to time for the Canadian longitudinal panel.

As can be seen from the statistical analysis and figures presented above, significant changes in user perceptions and intent within specific disclosure scenarios for the Canadian longitudinal panel occurred during the period of this study. More specifically, the data and statistical analyses indicate that significant movement first started to occur between the months of December of 2009 and May of 2010.

The implications of this movement will be examined in the implications sections of this paper.

Given the dramatic shifts in many of the variates tested in the above iterations, questions arose about why these shifts occurred. Consequently, during the month of February 2011, I conducted a series of six qualitative interviews with selected Canadian Longitudinal Panel respondents and American respondents to ascertain the motivations and perceptions that produced these changes in attitudes and reported behaviors.

These qualitative interviews followed a loosely structured format that asked the following series of open-ended response questions:

**Table 6 – Qualitative Interview Questions**

| | Interview Questions |
|---|---|
| 1 | Why have you reduced your disclosure of personal information on Social Networking Sites during the past two years? |
| 2 | Why have you increased your falsification of personal information on Social Networking Sites during the past two years? |
| 3 | Why has your distrust of social networking sites increased during the past two years? |
| 4 | Why has your sense of risk from Social Networking Sites increased during the past two years? |
| 5 | Why have you reduced your willingness to friend strangers in the past two years? |
| 6 | Do you regard some Social Networking Sites as being 'riskier' than others? |
| 7 | If you had a way of monitoring your level of personal disclosure risk, would you use it? |

The purpose of these qualitative interviews was to establish individual's stated reasons behind their change in privacy behaviors, and also to ascertain what differences and similarities existed in perceptions between Canadian and American respondents. While American respondents varied over iterations of the study, some respondents remained consistent and similar tightening of privacy attitudes was noted between the Canadian and American respondents. If their stated reasons for these changes were similar, some support is provided for the claim that this phenomenon is common to both countries.

**Respondent Profiles**
Respondents were selected on the basis of identified changes in privacy attitudes and stated behaviors, demographic diversity and availability for in-person or telephone interviews.

Given that these respondents had been tracked over multiple iterations of the study survey, and had been identified as representative of their particular demographic profiles, I already knew that their reported attitudes and behaviors had changed over time and that these changes were statistically significant within a confidence value of 0.99. Thus, it was not necessary to ask if these reported attitudes and behaviors had changed, only to ask why they had changed.

**Qualitative Interview Responses**
Interview respondents displayed a marked degree of uniformity in their range of answers, however, the reasons stated for specific behaviors varied from question to question. For example, for different respondents, their dislike of targeted online marketing was stated as answers to different questions. However, a few key global response trends were readily apparent in all responses.

*Increased familiarity with Social Networking Sites:* A number of respondents reported changes in their privacy attitudes and behaviors as a result of increased familiarity with Social Networking Sites. In the words of a Canadian respondent "I know a lot more (about social networking) than I did a couple of years ago. Implicit within these answers was the recognition that as respondent's familiarity with social networking increased, their privacy behaviors became better informed.

*Increased awareness and recognition of others' maladaptive behaviors:* The most common reason, stated in a variety of ways by respondents across the Canadian and American groups was increased knowledge and awareness of online privacy risks. This increased knowledge has arisen from media exposure, conversations and input from friends and family and—perhaps most significantly for the majority of respondents—by observing inappropriate online privacy behaviors by their social networking peer group. In the words of one respondent: "I would never post some of the things I have seen posted."

*Dislike of targeted online marketing campaigns:* "Customized ads are just plain creepy." stated one American respondent. In the words of a Canadian respondent: "When I found out they were selling my data, I shut down a lot of my activity." While these two statements may not seem directly related, a strong theme emerged from respondents of awareness that their personal data was being used for targeted marketing purposes. These responses imply two increases in awareness.

All respondents volunteered their awareness that their personal data was being collected and used for commercial purposes. However, no respondents ceased their social networking activities as a result of this awareness. Instead, information disclosure reduction, information falsification, reduction in 'friending', and increasingly cautious behavior were adopted by respondents as adaptive strategies.

*Potential for information misuse by stalkers:* One significant difference was noted among respondents. All female respondents stated awareness of the risk of personal information misuse by stalkers, while only one male respondent identified this risk, and then only in regards to his children. It is possible that this difference in responses between genders may be correlated to the slightly more conservative privacy attitudes and reported behaviors identified for women in Iteration 1 of this study.

*Reduced willingness to 'friend' requests from unknown individuals:* Different combinations of the above factors were cited by various respondents as reasons for reducing their willingness to 'friend'. As such, decreased willingness to 'friend' was a result of perceptions and learning on the part of the respondents, rather than a causal factor in this learning. However, the different emphasis each respondent placed on factors underlying their decisions to 'friend' suggest that each user constructed assigned unique weightings to each of the above inputs and that each respondent developed unique strategies to cope to mitigate their perception of increased risks.

This dynamic re-evaluation of privacy boundaries by users is based upon a dialectical process of observation of context, engagement in that context, and learning through experience is central to the boundary regulation model presented in this study and is quantitatively validated in the study results presented in the following chapter.

## 5. Study Limitations

A longitudinal study of this nature has a number of inherent limitations. The first--and most obvious--is temporality. This study would ideally occur over a period of decades, rather than a period of two years. However, these have arguably been the most dynamic twenty four months in the history of online privacy, and constitute the best temporal opportunity to date for a study of this nature. The second limitation arises from the manner in which study participants were recruited throughout successive iterations of this study.

Within a Longitudinal Panel Study, a demographic cross-section of respondents is tracked over a period of time. While the consistency of this study's

Canadian solicitation pool was maintained across successive iterations of the study, there was no way to ensure the consistency of response across these iterations. For example, approximately 90% of Canadians solicited by the author responded to each iteration of the study, but there is no way of determining whether it was the same 90% each time. However, given the high rates of Canadian response across iterations of the study, and demographic consistency among Canadian longitudinal panel respondents across iterations, the validity of this longitudinal panel can be asserted.

Further complicating the limitations of these results are internal validity factors common to many longitudinal studies.

*Repeated Testing Sensitization:* By virtue of having participated in a privacy study four times over a period of two years, Canadian study participants may have become sensitized to privacy concerns and this sensitization may have affected their privacy attitudes and behaviors. However, given the level of online privacy hysteria evidenced in the popular press during that past twenty-four months, it would be difficult for any person, participant or not, to avoid some sensitization on this issue. Additionally, the same statistically highly significant tightening of privacy attitudes and behaviors can be detected in the Pace University community responses that were not drawn from a longitudinal panel. This suggests that these increasing concerns are common to many social media users, not just the ones tracked in the Canadian longitudinal panel.

*Confounding:* The causal relationships among variables examined in this study may, in fact, be due to factors not included within the conceptual model. For example, as stated above, the media has the capacity to shape opinion and perhaps behavior. Consequently, this study includes media exposure, input from family and friends, and personal observations as variables for inputs into privacy attitudes. Given the relatively new nature of this field, and the rapidity with which it is evolving, it is possible that influential factors exist that have not been captured in the model. However, the breadth of research that underpins this study's taxonomy and conceptual model helps ensure that the model is as inclusive as it can possibly be given the current state of scholarship.

*Self-Selecting Respondents:* Is a respondent to a privacy questionnaire solicitation likely to already be more sensitized to privacy concerns? This is difficult to determine. Given that this study examines changing perceptions over a period of time by measuring the amount of change, the 'entry level' of privacy attitude is not as important as it would be in a time invariant study. This self selecting limitation is probably more relevant to the Pace University respondents, who were a varying group drawn from a larger solicitation pool, rather than the Canadian respondents, who were a consistent group with a very high response rate.

*Self-Reporting of Behavior:* Any self-reported behavior is inherently suspect, and is inferior to clinical observation of the test participant. This study limitation should be acknowledged and study conclusions should be appropriately qualified as a result of this limitation.

*Breadth of Reach of Solicitation Pools:* This longitudinal study was, by its very nature, heavily biased to Canadian participants. However, the Canadian Longitudinal Panel was relatively ethnically diverse, with approximately fifty percent of respondents reporting Caucasian ethnicity.

This study addresses the above limitations in as coherent a manner as possible, and appropriately bounds the conclusions based on the limitations inherent in this study.

## 6. Implications

Given that a significant shift in respondent attitudes and reported behaviors occurred during the period of this study, one questions how and why these changes occurred. The most obvious explanation is that, during this period users became more aware of social media privacy risks and adjusted their behavior accordingly. However, this study also tested for media exposure to privacy risks and did not find a statistically significant shift in responses. During the period of this study, most major media outlets carried at least one online privacy risk story every week and most respondents reported being very aware of media coverage of online privacy threats throughout the entire study. Consequently, the shift in media sensitization during the period of this study was not statistically significant, with a Chi-square probability of 0.77 between Iteration 1 and Iteration 4.

Another possible explanation for the observed tightening of privacy attitudes and reported behaviors is respondents perceiving themselves to be victims of social media privacy violation. The study also tested for that variable, however, throughout the course of the study, most respondents did not feel themselves to have been a victim of privacy violation. With a Chi-square probability of 0.63 between Iteration 1 and Iteration 4, this variable did not change in a statistically significant manner.

Additionally, one would expect users who are increasing privacy concerns and lessening reported disclosure to place lesser importance on having a large social network, due to the increased privacy risks presented by a larger social network. This was not the case. During the period of this study, respondents reported the increasing importance with which they viewed social network size. This change was highly significant, with a Chi-square probability of 3.1E-20 between Iteration 1 and Iteration 4. A diagrammatic representation of the above factors appears below. As with the previous graphs, the y-axis denotes the means of responses on a 7 point Likert Scale where 1="Strongly Agree" and 7="Strongly Disagree".

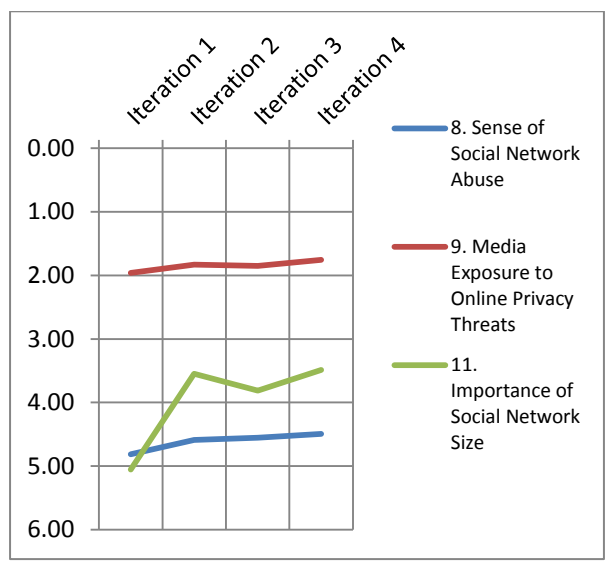

**Figure 4–Additional Privacy Disclosure Variables**

The data above suggests that a number of complex underlying factors are influencing online privacy attitudes and behaviors and that these behaviors appear to be, in some cases, counterintuitive. While user privacy concerns are increasing and reported disclosure is decreasing, media exposure and sense of victimization are remaining relatively consistent. Additionally, importance of social network size is growing.

It would appear that a number of these variables are not correlated, and are operating independently of each other in an unknown fashion. This also suggests a complex set of underlying drivers for these attitudes and behaviors.

## 7. Future Directions for Research

The most fertile field of research suggested by these findings is the creation of a model that can help explain these seemingly paradoxical movements in variables.

By expanding the range of input variables considered by online users as they make privacy disclosure decisions, it may become possible to test for a multivariate set of factors that actually drive user decisions. This model creation and validation was the subject of a recent study by this author and results will be presented in an upcoming paper.

It is probable that this dissertation has taken this research as far as it can with existing survey instruments. Rather than constituting an academic dead end, this suggests possibilities for a number of future investigations.

*Longitudinal Extension:* At the most basic level, the validated construct within this study could be extended and administered over the coming months and years to evaluate continued changes in privacy attitudes and behaviors.

*Laboratory Study:* Research opportunities exist for closer examination of how individuals make specific privacy choices, and what inputs influence these choices. A 'live' laboratory study provides the researcher with opportunities to examine choice and behavior that survey instruments do not present.

*Educational Study:* In a similar manner, given that this dissertation has shown that user perceptions and experience form crucial inputs into specific privacy disclosure choices, the impact and value of privacy education on user disclosure behavior should be examined.

*Instrumentation Study:* Given that this dissertation has shown that specific privacy choices by individuals are based upon a multiplicity of factors, the question arises as to whether software instrumentation might help assist users in the execution of complex choices that are dependent on a wide range of variables.